\documentclass[apj]{emulateapj}

\usepackage{natbib}
\usepackage{graphicx}
\bibpunct{(}{)}{;}{a}{}{,}

\newcommand\modified[1]{{ #1}}

\begin{document}

   \title{
   On the use of the index N2 to derive the metallicity in metal-poor galaxies
}
   \author{
         	A.~B. Morales-Luis\altaffilmark{1,2},
          	E. P\'erez-Montero\altaffilmark{3},
          	J.~S\'anchez~Almeida\altaffilmark{1,2},
                \and
                C.~Mu\~noz-Tu\~n\'on\altaffilmark{1,2}}

\altaffiltext{1}{Instituto de Astrof\'\i sica de Canarias, E-38205 La Laguna, Tenerife, Spain}
\altaffiltext{2}{Departamento de Astrof\'\i sica, Universidad de La Laguna, Tenerife, Spain}
\altaffiltext{3}{Instituto de Astrof\'\i sica de Andaluc\'\i a - CSIC, Apdo. de correos 3004, E-18080 Granada, Spain}

\email{abml_ext@iac.es,epm@iaa.es,jos@iac.es,cmt@iac.es}
\begin{abstract}
The N2 index ([N{\sc ii}]$\rm\lambda$6584/H$\rm\alpha$)
is used to determine emission line galaxy metallicities at all redshifts, including
high redshift, where galaxies tend to be metal-poor. 
The initial aim of the work was to improve the calibrations used
to infer oxygen abundance from N2 employing updated
low-metallicity galaxy databases. We compare  
N2 and the metallicity determined using the direct method for the set of extremely metal-poor
galaxies compiled by \citet{2011ApJ...743...77M}.
To our surprise, the oxygen abundance presents a tendency to be constant with N2, with a very large scatter. Consequently,
we find that the existing N2 calibrators overstimate the oxygen abundance for most low metallicity 
galaxies, and then they can be used only to set upper limits to the true metallicity in low-metallicity galaxies. 
An explicit expression for this limit is given. In addition, we try to
explain the observed scatter using photoionization models.
It is mostly due to the different evolutionary state of the H{\sc ii} regions producing the
emission lines, but it also arises due to differences of N/O among the galaxies.
\\
\vskip 3cm 
\end{abstract}




\section{Introduction}
\label{sec:intro}

Optical bright emission lines have been used for many
years as the main source of information to derive physical properties and chemical
abundances both in nearby and distant galaxies. These optical emission-lines are 
excited by the UV radiation emitted by 
young massive stars in the gas clouds surrounding
on-going star-formation complexes in galaxies.
The total metallicity ($\rm Z$) of the gas-phase is one of the most relevant 
pieces of information that can be extracted.
In particular, the regime of low metallicity is particularly important in the context of
unevolved/young objects resembling the first protogalaxies
\citep[e.g.,][]{2006A&A...448..955I,2011ApJ...743...77M} and or star-formation in the high redshift
universe \citep[e.g.,][]{2010Natur.467..811C,2011MNRAS.414.1263M,2014A&A...563A..58T}.

The most accurate method to derive $\rm Z$ using the bright collisionally excited emission-lines 
is the determination of the total oxygen abundance\footnote{O abundance is used as a
proxy for $\rm Z$ assuming, e.g., that the different metals are present in solar proportion. In addition, 
the nebular oxygen abundance inferred from emission lines is not the total O since part
of the O may be depleted into dust grains. However, this depletion is negligible small in our
galaxies as discussed at the end of Section~\ref{sec:physical_porperties}. Therefore in this paper
we use the term oxygen abundance even though we measure nebular oxygen abundance.}; \modified{O/H}
based on measuring the electron 
temperature, the so-called $\rm T_{e}$ method.
$\rm T_{e}$ can be derived from the ratio of lines 
with different excitation potential; the most widely used is the ratio of
oxygen lines, [O{\sc iii}]$\lambda$4363/([O{\sc iii}]$\lambda4959 +$ [O{\sc iii}]$\lambda5007$). 
Then, the ionic abundances of the most abundant
oxygen ions is obtained from the ratio of their brightest emission-lines to a 
Hydrogen Balmer line, {\rm i.e.,} [O{\sc ii}]$\lambda$3727/H$\beta$ for O$^+$ and
[O{\sc iii}]$\lambda$5007/H$\beta$ for O$^{2+}$ \citep[e.g.,][]{1992MNRAS.255..325P,2008MNRAS.383..209H}.

Unfortunately, the  $\rm T_{e}$ method is often useless 
at large redshifts due to the faintness of the weak
auroral lines and/or because the wavelength coverage  
does not include the emission-lines required to derive of the observation electron temperature
or ionic abundances.
In those cases, other methods based only on the
brightest available emission-lines are used instead. One of the most 
popular ones is based on the ratio between [N{\sc ii}]$\lambda$6584 and
H$\alpha$, the so-called N2 parameter \citep[e.g.,][]{1994ApJ...429..572S,2002MNRAS.330...69D,2004MNRAS.348L..59P,
2009MNRAS.398..949P}. This parameter
involves emission lines close in wavelength so it 
is almost independent of reddening or flux calibration uncertainties. N2 presents
a linear relation with \modified{log(O/H)}, within a range of \modified{metallicity}
including extremely metal poor (XMP) galaxies\footnote{By definition, galaxies with
$\rm Z<Z_{\odot}/10$; see, e.g., \citet{2000A&ARv..10....1K}.} 
\citep[e.g.,][]{2002MNRAS.330...69D,2004MNRAS.348L..59P, 2009MNRAS.398..949P,
2012MNRAS.419.1051L,2013A&A...559A.114M}. Moreover, N2 allows measuring
\modified{O/H} in high redshift objects, because the required [N{\sc ii}] emission-line appears 
in many high-z surveys focused on the detection of H$\alpha$ in
near-IR bands \citep[e.g.,][]{2008ApJ...674..151E,2009A&A...506..681Q,2012A&A...539A..93Q}.
Other methods widely used to derive \modified{O/H} based on 
strong collisional emission lines have revealed to be quite inefficient
in the range of XMPs. For instance, the R23 parameter \citep{1979MNRAS.189...95P} has a bi-valuated
behaviour with \modified{O/H} in high redshift objects, and O3N2 \citep{1979A&A....78..200A}
cannot be used for oxygen abundance 12+log(O/H)$<$8 \citep[e.g.,][]{2013A&A...559A.114M}.

This paper is focused on the empirical \modified{O/H} estimate using
the N2 index in metal-poor galaxies. We wanted to take advantage of the comprehensive sample of metal-poor galaxies
compiled by \citet{2011ApJ...743...77M} to improve the
significance of the  calibrations \modified{O/H} versus N2 obtained
so far \citep[e.g.][]{2002MNRAS.330...69D, 2004MNRAS.348L..59P,2006A&A...459...85N,
2009MNRAS.398..949P}. To our surprise, we
found that the scatter of the relationship \modified{O/H} versus N2 increases when improving the statistics, and reveals 
that the ratio [N{\sc ii}]$\lambda$6583 to H$\alpha$ seems to be independent of metallicity at low oxygen 
abundance (12+log(O/H)$<$7.6, which corresponds to a metallicity $\rm Z\lesssim Z_{\odot}/10$).
This result casts doubts on the metallicities 
of high-redshift metal-poor objects based on N2, but indicates that N2 can be used to set an upper limit
to the true metallicity of the targets. 
This paper describes the problem, the solution, and explains how the behaviour of N2 can be understood using 
photoionization models.

The paper is organized as follows. First, we describe the sample of 46 XMP galaxies 
(Section~\ref{sec:sample}) used to calibrate the N2-based empirical metallicity estimate.
The $\rm T_{e}$ method we use to determine physical properties 
and chemical abundances is presented in Section~\ref{sec:physical_porperties}.
The $\rm T_{e}$ method is employed to determine the chemical abundances of the sample of XMPs in
Section~\ref{sec:metallicity}.
In order to identify the sources of scatter in the N2 calibration at low metallicities,
photoionization models are analyzed in Section~\ref{sec:models}. The
sources of the observed scatter
are identified in Section~\ref{sec:results_empiric}. A summary with conclusions and 
follow-up work is provided in Section~\ref{sec:discussion}.

\bigskip
\section{Extremely Metal Poor galaxy sample}\label{sec:sample}

In order to calibrate the index N2 in the very low-Z range, 
we need a large sample of low-Z targets with their \modified{O/H}
homogeneously determined via the direct method.

We start off from the metal-poor galaxy sample compiled by \citet{2011ApJ...743...77M}, which 
included all galaxies with metallicity 
one tenth or less of the solar value \citep[i.e., 12 + log(O/H)$<$7.69,][]{2009ARA&A..47..481A} 
found to the date of publication.
The total sample includes 140 XMPs. Among them, 79 have spectra in SDSS/DR7, which is the spectral
database used in the calibration.

Nearby galaxies with spectra in SDSS 
(objects with $\rm redshift\lesssim 0.024$) have [O{\sc ii}]$\lambda$3727 out
of the observed spectral range, thus the determination of the $\rm O^{+}/H^{+}$ by the 
direct method, i.e., using [O{\sc ii}]$\lambda$3727, is not possible. 
A slight modification of the direct method allows to calculate
the value of $\rm O^{+}/H^{+}$ from the intensities of the auroral lines 
[O{\sc ii}]$\lambda\lambda$7320,7330 \citep{1984ASSL..112.....A} and was explored.
It is important, however, to bear in mind that the application of the auroral line method is restricted
to spectra with sufficiently high signal-to-noise ratio \citep{2004ApJS..153..429K}. Because of this,
we select galaxies with S/N $\gtrsim$ 3 in [O{\sc ii}]$\lambda$7320.
Only 31 of the 79 metal-poor galaxies with SDSS spectra fulfill the required S/N criterion.
We added 15 galaxies to the sample taking into account the 
similarity between N and O ionization structures. With N/O obtained using the 
N2S2 index \citep{2009MNRAS.398..949P} and $\rm N^{+}/H^{+}$ inferred through the 
functional form by \citet{2008MNRAS.383..209H}, it is possible to estimate $\rm O^{+}/H^{+}$
when [O{\sc ii}] lines are unavailable. In summary,
we have 46 XMP galaxies with their metallicities determined in a  
homogeneous and consistent way, which is the main sample employed in this paper.

In order to put our work into context, we also randomly select and 
analyze a control sample of 65 starburst galaxies, having intermediate 
metallicities ($\rm 7.7\leq12+log(O/H)\leq8.2$).
We first select those emission line galaxies in SDSS/DR7
which were not classified as metal-poor galaxies by \citet{2011ApJ...743...77M}.
Then, we discarded AGNs using the BPT diagram \citep{1981PASP...93....5B}. Finally, we divided the range
of N2 roughly corresponding to $\rm 7.7\leq12+log(O/H)\leq8.2$ in 65 equal intervals, and then one galaxy was
randomly selected per interval.

 \section{Technique to derive the physical properties and chemical abundances}\label{sec:physical_porperties}

The emission-line fluxes of the spectra of the 46 XMPs and the 65 galaxies 
in the control sample are measured using our 
own IDL procedure to better control the errors and their propagation.
Basically, the procedure works as the task SPLOT
in IRAF\footnote{IRAF is the Image Reduction and Analysis Facility distributed by
the National Optical Astronomy Observatory, which is operated by the
Association of Universities for Research in Astronomy (AURA) under
cooperative agreement with the National Science Foundation (NSF).
}.
In the case of an isolated line or two blended unresolved
lines, they are measured integrating between two points given by the 
position of the local continuum identified by eye. If two lines are 
blended, but they can be resolved, we use a multiple Gaussian fit 
to estimate individual fluxes.

The statistical error associated with the observed emission lines 
is calculated using the expression 
$\rm \sigma_{1}=\sigma_{c}N^{1/2}[1+EW/(N\Delta)]^{1/2}$ 
\citep{1994ApJ...437..239G,2002MNRAS.337..540C,2003MNRAS.346..105P},
where $\rm \sigma_{1}$ is the error in the observed line flux, $\rm \sigma_{c}$ 
corresponds to the noise in the continuum
near the measured emission line, $\rm N$ 
is the number of pixels used in the 
measurement of the line flux, $\rm EW$ is the line equivalent
width, and $\rm \Delta$ is the wavelength dispersion.

All line fluxes were corrected for reddening using 
the same procedures as in \citet{2008MNRAS.383..209H}.
The reddening coefficient in H$\rm \beta$, $\rm c(H\beta)$, was calculated assuming 
the extinction law of \citet{1972ApJ...172..593M} for the Galaxy   
and performing a least-squares fit to the 
difference between the theoretical and observed
Balmer decrement. The corrected emission line flux of a line with wavelength $\rm \lambda$, 
$\rm I(\lambda)$, is given by
\begin{equation}
 \rm\frac{I(\lambda)}{I(H\beta)}=\frac{F(\lambda)}{F(H\beta)}\cdot10^{c(H\beta)f(\lambda)},
\end{equation}
where $\rm F(\lambda)$ stands for the measured emission line flux, 
and $\rm f(\lambda)$ is the extinction law at the corresponding
wavelength.

With the reddening corrected fluxes, we determine the physical conditions of  the ionized
gas, electron temperature and electron density included. 
Electron temperature and electron density determinations are
based on the five-level statistical equilibrium atom approximation in the task TEMDEN of IRAF 
\citep{1987JRASC..81..195D}.
Electron density is derived  from the ratio [S{\sc ii}]$\lambda$6717/[S{\sc ii}]$\lambda$6731, and
in objects without this ratio, a density of 100 $\rm cm^{-3}$ is assumed, typical in this kind of galaxies
\citep[e.g.][]{2005MNRAS.361.1063P}. The electron temperature of [O{\sc iii}] is derived using the ratio 
([O{\sc iii}]$\lambda$4959+[O{\sc iii}]$\lambda$5007)/[O{\sc iii}]$\lambda$4363
\citep[e.g.,][]{2004cmpe.conf..115S}.

As we mentioned in Sec.~\ref{sec:sample}, [O{\sc ii}]$\lambda$3727 is often out of 
the wavelength range in the SDSS spectra
of our galaxies, thus the electron temperature of [O{\sc ii}] is calculated using the model relation between 
t([O{\sc ii}]) and t([O{\sc iii}]) worked out by \citet{2003MNRAS.346..105P}. 
Furthermore, in the 15 cases without
$\rm O^{+}/H^{+}$, we need
the electron temperature of [N{\sc ii}] to calculate $\rm N^{+}/H^{+}$.
t[N{\sc ii}]) can be derived from the emission-line ratio
([N{\sc ii}]$\lambda$6584 +[N{\sc ii}]$\lambda$6548)/[N{\sc ii}]$\lambda$5755 in 8 objects.
The remaining 7 cases use
an independent relation between t([N{\sc ii}]) and t([O{\sc iii}]) 
worked out by \citet{2009MNRAS.398..949P}.

This particular study uses oxygen abundance to trace the metallicity, therefore, 
we have not calculated the abundance of other elements.
The measures calculates the total oxygen abundance relative 
to hydrogen using, 
\begin{equation}
\rm\frac{O}{H}\backsimeq\frac{O^{+}+O^{2+}}{H^{+}}.
\label{eq:oxabexpression}
\end{equation}
The ionic abundances of $\rm O^{+}$ and $\rm O^{2+}$ are obtained employing the expressions given by
\citet{2008MNRAS.383..209H}, which consider the lines 
[O{\sc iii}]$\rm\lambda$4959 and [O{\sc iii}]$\rm\lambda$5007 for $\rm O^{2+}$, and 
[O{\sc ii}]$\rm\lambda\lambda$7320,7330 for $\rm O^{+}$.
O$^{\rm3+}$ does not need to be included since its abundance is negligibly small in H{\sc ii} regions 
\citep[e.g.,][]{2012EAS....54....3S,2013ApJ...765..140A}.
For those galaxies without a direct determination of 
the ionic abundance $\rm O^{+}/H^{+}$, considering the similarity of ionization 
structures of N and O, it is possible to assume,
\begin{equation}
\rm\frac{N}{O}\backsimeq\frac{N^{+}}{O^{+}}.
\label{eq:n_o}
\end{equation}
With N/O obtained using the N2S2 index by \citet{2009MNRAS.398..949P}, and $\rm N^{+}/H^{+}$,
calculated employing the approximation by \citet{2008MNRAS.383..209H},
it is possible to determine $\rm O^{+}/H^{+}$ through the expression,

\begin{equation}
\rm\frac{O^{+}}{H^{+}}=\frac{N^{+}}{H^{+}}\frac{O}{N}.
 \label{eq:o_h}
\end{equation}

Errors in chemical abundances and physical properties are estimated in a Montecarlo simulation.
The line fluxes involved in determining
the physical conditions and chemical abundances are randomly modified 500 times according to 
their observed errors, and the abundances are computed for each one of these realizations. 
The error for the abundance is
the standard deviation among all the values thus obtained.
The noise added to the observed fluxes is assumed to be Gaussian. 
\modified{These error bars do not include systematic errors, which may be 
non-negligible. For instante, even XMP galaxias, where the
dust is almost absent \citep[e.g.,][]{2014Natur.505..186F},
may have 20\% of their O depleted into dust grains \citep[e.g.,][]{2010ApJ...724..791P}.
This represent
an artificial drop of metallicity of 0.08\,dex. Similar error results
from ignoring temperature inhomogenities \citep[e.g.,][]{2010ApJ...724..791P}.}

\begin{figure}[h!!!!]
  \centering
    \includegraphics[width=0.5\textwidth]{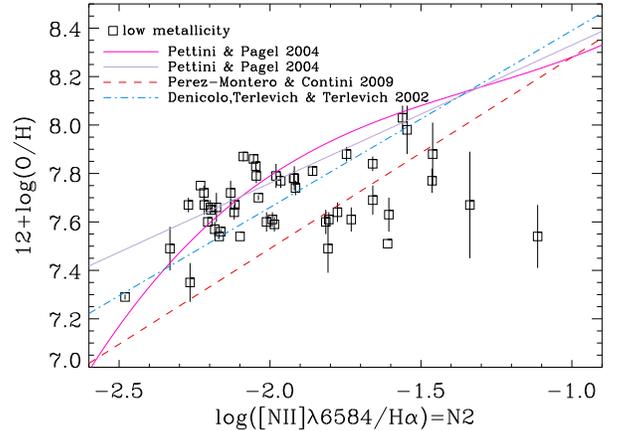}
    \caption{Oxygen abundance versus N2 for the XMP galaxy sample. 
    We also include the third-order polynomial calibration of the relationship from PP04 (the solid line), the linear fit calibration from
PP04 (the dash triple-dot line), the
linear calibration from \citet{2009MNRAS.398..949P} (the dashed-line) and the
linear calibration from \citet{2002MNRAS.330...69D} (the dash-dot line).}
  \label{fig:results_metalpoor}
\end{figure}

     \begin{figure}
 \centering
    \includegraphics[width=0.5\textwidth]{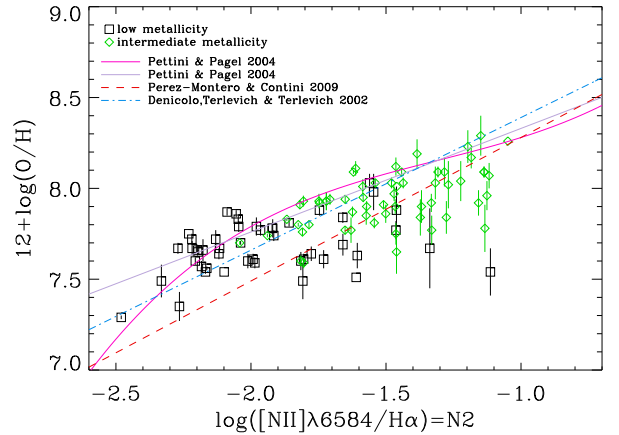}
  \caption{Oxygen abundance versus N2 for the full sample, i.e., the low metallicity
  sample (the black squares) plus the control sample (the green diamonds). We also include
  the third-order polynomial calibration from PP04 (the solid line), the linear fit calibration from
PP04 (the dash triple-dot line), the
linear calibration from \citet{2009MNRAS.398..949P} (the dashed-line) and the
linear calibration from \citet{2002MNRAS.330...69D} (the dash-dot line). The lines represent typical calibrations used 
in the literature to derive O/H from N2.}
  \label{fig:results_allsample}

\end{figure}

\smallskip

\section{Metallicity and N2 index}\label{sec:metallicity}

The \modified{oxygen nebular metallicity} computed for the metal-poor sample is compared with N2 in
Fig.~\ref{fig:results_metalpoor}. The first important result is that, most of the galaxies
have metallicities one tenth or less of the solar value. Therefore, the oxygen abundance obtained 
using the direct method confirms that most
galaxies selected as low-metallicity galaxies in \citet{2011ApJ...743...77M} are, indeed, XMPs.

In addition, the observed oxygen abundance
presents a tendency to be constant with N2 for the metal-poor galaxies, 
i.e., metallicities have similar values
for a broad range of [N{\sc ii}]$\lambda6583$/H$\alpha$ (Fig.~\ref{fig:results_metalpoor}).
This is discouraging if one tries to use N2 as proxy for metallicity at low metallicity. As
we argue below, N2-based metallicity estimates by
\citet[][PP04]{2004MNRAS.348L..59P} calibrate the N2 index 
versus \modified{O/H} relationship in the range $\rm 7\leq12+log(O/H)\leq8.7$ 
(see Fig.~\ref{fig:results_metalpoor}).
The data set used by PP04 presents a discontinuity in metallicity in the region of XMP galaxies.
Figure~\ref{fig:results_allsample} shows 
our metal-poor galaxy sample together with the control
sample. There is no jump in the low metallicity range. 
Then, we conclude that the apparent discontinuity in PP04 is due 
to the small number of data points they had available. The present sample makes use of a
more recent and larger database and, the oxygen abundance versus N2 shows a
continuous trend even in the low metallicity range.

The scatter of the relationship $\rm\log(O/H)$ vs N2 is very large (Fig.~\ref{fig:results_allsample}).
\modified{This scatter is also present in other samples like the one used by \citet{2004MNRAS.348L..59P}
and the more recent sample analyzed by \citet{2012ApJ...754...98B}.}
We initially attempted to 
calibrate the N2 index in the low metallicity range, but we gave up due to the scatter of the observed
points. However, N2 can be safely used to set an upper limit
to the true metallicity in the low metallicity range.
The third-order polynomial calibration from PP04 suffices for this objective
(see the fit inserted in Fig.~\ref{fig:results_allsample}), namely,
\begin{equation}
 \rm 12+log(O/H)\lesssim 9.37+2.03\cdot N2+1.26\cdot N2^{2}+0.32\cdot N2^{3},
\label{eq:poli3_pp04}
 \end{equation}
when $\rm -2.5\leq N2\leq-1$. Equation~\ref{eq:poli3_pp04} grants that a XMP galaxy according to N2 is a truly XMP galaxy.

\section{Origin of the scatter in the metallicity versus N2 relationship}

\subsection{Photoionization models}\label{sec:models}

In order to understand the source of scatter in Figs.~\ref{fig:results_metalpoor} 
and \ref{fig:results_allsample},
we developed a set of photoionization models using CLOUDY \citep[v.10; ][]{1998PASP..110..761F},
covering the physical condition expected for the observed objects. 

The photoionization models assume a spherically symmetric H{\sc ii} region, with the ionized 
emitting gas taken to have a constant density of 50 $\rm cm^{-3}$. 
\modified{We also tried models with constant gas pressure, but they do not 
significantly modify the emission line ratios with respect to constant density
models, and so this assumption does not bias the conclusions of the modeling.} 
The mo\-dels have plane-parallel geometry.
The gas is ionized by a coeval cluster of massive stars, 
with the spectral energy
distribution (SED) obtained using Starburst99 synthetic stellar 
atmospheres \citep{1999ApJS..123....3L}. 
The age of the cluster is 1 Myr. 
We use an initial
mass function (IMF) with exponents 1.3 and 2.3 at low and high masses, respectively. Boundaries 
for the IMF are 0.1 and 100~$\rm M_{\odot}$, and the exponent changes at 0.5~$\rm M_{\odot}$.
Models use Padova AGB stellar tracks with
metallicities Z=0.004 
\modified{and Z=0.0004, the latter being the lowest available metallicity. Assuming solar 
composition \citep{2009ARA&A..47..481A}, Z=0.0004 $\equiv$ 12+log(O/H)=7.16 and 
Z=0.004 $\equiv$ 12+log(O/H)=8.16, which corresponds to the range of oxygen abundance found in our targets.}

\begin{figure}
\centering
\includegraphics[width=0.5\textwidth]{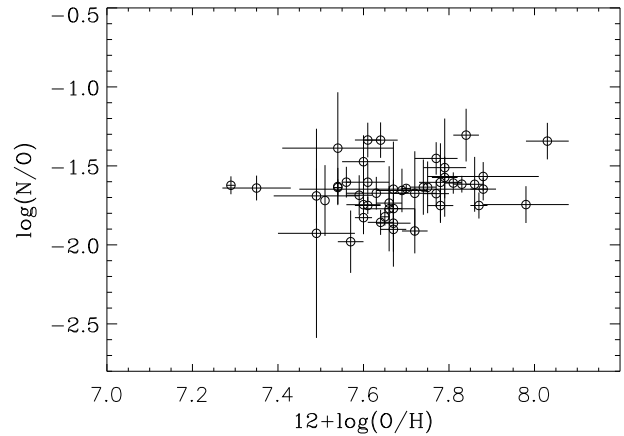} 
\caption{log(N/O) versus 12+log(O/H) for the metal-poor galaxy sample using N2S2
from \citet{2009MNRAS.398..949P}. Error bars are calculated as is explained in Sec.~\ref{sec:physical_porperties}.}
\label{Fig:N_O}
\end{figure}
\bigskip

The available metallicities are too coarse for the intended modeling. Therefore, 
for intermediate metallicity, we interpolate the two extreme spectra, one with $\rm Z_{1}=0.0004$ and 
the other with $\rm Z_{2}=0.004$. If
\begin{equation}
 \rm Z=w_{1}\cdot Z_{1}+w_{2}\cdot Z_{2},
\end{equation}
then the spectrum corresponding to metallicity Z, $\rm S_{Z}$, is
\begin{equation}
 \rm S_{Z}=w_{1}\cdot S_{Z_{1}}+w_{2}\cdot S_{Z_{2}},
\end{equation}
with $0 < \rm w_{1},w_{2} < 1$ and, $\rm w_{1}+w_{2}=1$.

It is also assumed that the gas has the same metallicity as the ionizing stars. The other ionic
abundances are set in solar proportions \citep{2009ARA&A..47..481A}, except in the case of nitrogen, where 
we explore different values of N/O.  Our galaxies are expected to have 
$\log (\rm N/O)\approx-1.6$ \citep{2000ApJ...541..660H}. To measure N/O one would ideally like to have
[O{\sc ii}]$\lambda$3727, 
but, as it is explained in Sec.~\ref{sec:sample}, the line is often outside the observed spectral range. 
In this case we use N2S2 index  
to obtain an approximate value of N/O \citep{2009MNRAS.398..949P}. Fig.~\ref{Fig:N_O} 
shows the result for our sample. From the figure we see 
that the metal-poor galaxies have typically $\rm -1.6\leq\log (N/O)\leq-1.4$.
Taking into account this fact, we compute two sets of models with $\log (\rm N/O)=-1.6$ and $\rm-1.4$. 
Each one is built for a number of ionization parameters $\rm \log U=-3,-2.9,-2.7,-2.5,-2.4,-2.2,-2$ 
and for a number of oxygen abundances $12+\log(\rm O/H)=7.16, 7.26, 7.36, 7.46. 7.56, 7.66, 
7.76, 7.86, 7.96, 8.06, 8.16$.  
The range of $\rm\log U$ corresponds
to the ionization degree shown by the
type of object studied here \citep[e.g.,][]{2005MNRAS.361.1063P}. All in all, we have 154 photoionization models
which allow us to predict N2 of a function of 12+log(O/H) under a large number of circumstances.

\subsection{Scatter in $\rm O/H$ vs $\rm N2$}\label{sec:results_empiric}

When looking for a metallicity
indicator based on bright emission lines such as N2, one would ideally like to find combinations of
lines whose fluxes depend only on chemical abundances. This is difficult because 
the emission line fluxes are controlled by other physical parameters as well.
In the case of N2, the ionization parameter (U), 
i.e., ratio between the number of ionizing photons and density of hydrogen atoms, is the parameter
cause most of the intrinsic variations \citep{2002MNRAS.330...69D}.

\begin{figure}
\centering
\includegraphics[width=0.5\textwidth]{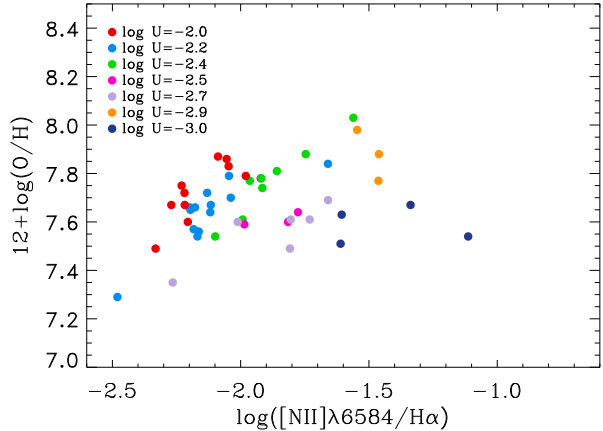} 

\caption{Oxygen abundance versus N2 for the metal-poor galaxy sample. Each color
is associated with one value of $\rm \log U$ as coded in the inset.
There is a clear trend for smaller N2 to have lower U. The 
relationship is not one-to-one, though.} 
\label{Fig:N2_sepU}
\end{figure}

In order to understand and identify the relationship between the scatter observed 
in Fig.~\ref{fig:results_metalpoor}
and the ionization parameter, we employ the photoionization 
models described in Sec.~\ref{sec:models}.  
[O{\sc iii}]/H$\beta$=[O{\sc iii}]$\rm(\lambda 4959 +\lambda 5007)/H\beta$  weakly
depends on metallicity in a non-trivial way, although it is mainly sensitive
to the ionization parameter at sub-solar metallicity 
\citep[e.g.,][]{1981PASP...93....5B,2001ApJ...556..121K,2010AJ....139..712L}. 
We use the observed [O{\sc iii}]/H$\beta$ together with the models to estimate U in individual galaxies.
Metallicity and 
N/O are known. Thereby, we plot [O{\sc iii}]/H$\beta$ versus N2 
for the models and the galaxies, 
and we assign to each galaxy the nearest value of U. 
The results are shown in Fig.~\ref{Fig:N2_sepU}, which contains
oxygen abundance versus N2 with different colors representing different ionization parameters. 
This plot is quite
revealing, because it indicates that the scatter observed in Fig.~\ref{fig:results_metalpoor}
can be explained by the galaxies having different degree of
ionization. As Fig.~\ref{Fig:N2_sepU} shows, given an oxygen abundance,
N2 decreases with increasing ionization parameter.

\begin{figure}
\centering
\includegraphics[width=0.5\textwidth]{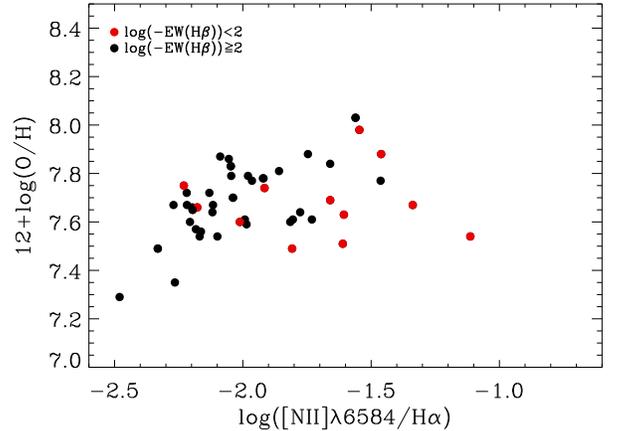} 

\caption{Oxygen abundance versus N2 for the metal-poor galaxy sample.
Red filled circles represent galaxies with EW(H$\beta$)$<$100\AA{}, whereas black filled circles 
stand for galaxies with EW(H$\beta$)$\geq$100\AA{}.}
\label{Fig:N2_sepEWHb}
\end{figure}

The degree of ionization of a galaxy changes with its evolutionary state 
\citep[e.g.,][]{2010AJ....139..712L}.
The observed equivalent width of H$\beta$ (EW(H$\beta$)) indicates 
the age of the ionizing cluster \citep[e.g.,][]{1999ApJS..123....3L}, although it also
depends on the underlying stellar component of the galaxies that provides most of the photons
in continuum wavelengths. \citet{2010MNRAS.404.2037P} showed that the underlying
population contribution is less than 10\% in a sample of H{\sc ii} galaxies 
similar to our targets. Therefore, we can use $\rm EW(H\beta)$ as a qualitative estimate
of the evolutionary state of our XMP sample. 
We computed EW(H$\beta$) for our galaxies, and the maximum
of the distribution is at EW(H$\beta$)$\simeq$100\AA{}. Galaxies in two different age ranges
are shown in Fig.~\ref{Fig:N2_sepEWHb}; the red filled circles correspond to EW(H$\beta$)$<$100\AA{} whereas 
the galaxies with EW(H$\beta$)$\geq$100\AA{} are shown as black filled circles.
Lower EW(H$\beta$)s correspond to younger galaxies.
Comparing Fig.~\ref{Fig:N2_sepU} and \ref{Fig:N2_sepEWHb}, we can see that the galaxies 
with low ionization parameter are the most evolved ones. 
These results indicate that part of the dispersion showed in 
Fig.~\ref{fig:results_metalpoor} is related to the evolutionary state of the galaxies through
the ionization parameter U.

\begin{figure}
\centering
\includegraphics[width=0.5\textwidth]{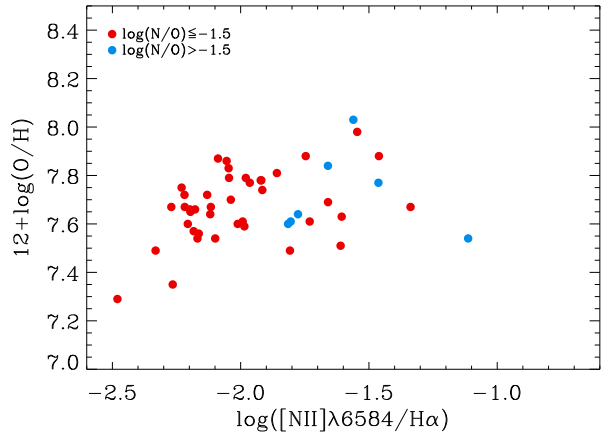} 

\caption{Oxygen abundance versus N2 for the metal-poor galaxy sample.
Red filled circles stand for galaxies log(N/O)$\leq$-1.5, whereas blue filled circles represent galaxies with
log(N/O)$>$-1.5.}
\label{Fig:N2_sepNO}
\end{figure}

The dependence of the scatter in Fig.~\ref{fig:results_metalpoor} on N/O was also explored. Since N2 uses
N line to measure O abundance, ti most depend on N/O.
As one can hinted at differences in Fig.~\ref{Fig:N2_sepNO}, another source of 
scatter seems to be in the N/O ratio.
The figure separates galaxies with log (N/O)$\leq$-1.5 (red filled circles) 
and galaxies with log (N/O)$>$-1.5 (blue filled circles).
Contrarily to what is expected for low-metallicity 
galaxies evolving as a closedbox system \citep[e.g.,][]{1978MNRAS.185P..77E,
1979A&A....78..200A}, N/O is not constant.

Figure~\ref{Fig:N_O} 
shows N/O for our metal-poor galaxies. 
The vast majority have
log (N/O)$\approx$-1.6, which is the value expected for metal-poor galaxies.
\modified{This plateau is though to begin at $\rm log(O/H)\lesssim 7.7$
\citep[e.g.,][]{2012ApJ...754...98B}, and the actual value of $\rm log(N/O)$ may depend on the type of 
galaxy \citep[e.g.,][]{2006ApJ...636..214V,2014ApJ...786..155N}.} 
\modified{There are galaxies 
with an excess of N/O in Fig.~\ref{Fig:N_O}.}
Those are the ones with low metallicity but large N2 in Fig.~\ref{Fig:N2_sepEWHb}
Possible reasons for this excess could be, extra production of primary nitrogen, coming from low-metallicity
intermediate-mass stars \citep[e.g.,][]{2006MNRAS.372.1069M, 2006A&A...450..509G} or
Wolf-Rayet stars \citep{2011A&A...532A.141P,2013AdAst2013E..18P}, or a combination of inflows of 
metal-poor gas and outflows of enriched gas \citep{2010ApJ...715L.128A,2014ApJ...783...45S,2014ApJ...786..155N}.
Nitrogen enhancements in similar galaxies have also been reported by 
\citet[][]{2009AJ....137.5068L,2010A&A...517A..27M,2011MNRAS.411.2076L,
2012ApJ...749..185A,2013MNRAS.428...86J,2013MNRAS.432.2731K}.

 \section{Summary and Final Remarks}\label{sec:discussion}

Our initial aim  was improving the calibration used in the literature
to infer oxygen abundance from N2 for low-metallicity galaxies. In particular, we compare 
N2 and metallicity determined using the direct method in the set of XMP galaxies worked
out by \citet{2011ApJ...743...77M}.
The first result is that the oxygen abundances obtained using
the direct method confirm that galaxies classified as low-metallicity galaxies in
\citet{2011ApJ...743...77M} are, indeed, metal-poor.
The second result is more discouraging though. The observed 
oxygen abundance presents a tendency to be \modified{constant} with N2 making it difficult to work out
any calibration at low metallicity. Instead, we argue that the calibration O/H vs N2 by PP04 can be used
at low metallicity to set an upper limit to the true metallicity (see Eq.~\ref{eq:poli3_pp04}). 

O/H vs N2 presents a very large scatter (see Fig.~\ref{fig:results_metalpoor}). 
CLOUDY photoionization models allowed us to understand the scatter. We found that
N2 decreases with the ionization parameter for a given oxygen abundance. 
Considering that the degree of ionization is related to the evolutionary state of the starburst
ionizing the medium, we analyze
the state of evolution using the equivalent width of H$\beta$. This analysis indicates
that part of the dispersion observed is indeed due to the evolutionary state of the galaxies.
In addition, we found that part of the scatter is also due 
to an excess of N/O in some of the metal-poor galaxies. An excess could be due to an 
extra production of primary nitrogen galaxies, or other processes including
inflows of metal-poor gas 
(Sec.~\ref{sec:results_empiric}).

In short, we find the commonly used metallicity estimate based on N2 to be uncertain at low-metallicities. 
Fortunately, it is possible to use the N2 calibration to set an upper limit to the abundance in metal-poor galaxies.

\smallskip

%
\begin{acknowledgements}

We are thankful to Elena Terlevich, Roberto Terlevich and Jose V\'ilchez for insightful comments and discussions.
ABML thanks the hospitality of the IAA during her visit. 
EPM thanks also to projects 
PEX2011-FQM7058 and TIC114  {\em Galaxias y Cosmolog\'\i a} of the
 Junta de Andaluc\'\i a (Spain)"
This work has been funded by the Spanish MICIN project {\em Estallidos},
AYA~2010-21887-C04-04.  
We are members of the Consolider-Ingenio 2010 Program, grant 
MICINN CSD2006-00070: First Science with GTC.
Funding for the SDSS and SDSS-II has been provided by the Alfred P. Sloan
Foundation, the Participating Institutions, the National Science Foundation, the
U.S. Department of Energy, the National Aeronautics and Space Administration,
the Japanese Monbukagakusho, the Max Planck Society, and the Higher Education
Funding Council for England. The SDSS is managed by the Astrophysical Research 
Consortium for the Participating Institutions (for details,
see the SDSS web site at http://www.sdss.org/).

{\it Facilities:} \facility{Sloan (DR7, spectra)
}
\end{acknowledgements}

%


%

%

\end{document}